\documentclass[%
reprint,
amsmath,amssymb,
aps,
nofootinbib,
pre,
twocolumn
]{revtex4-1}

\usepackage{mathbbol}
\usepackage{graphicx}
\usepackage{dcolumn}
\usepackage{bm}

\newcommand{\la}[1]{\label{#1}}
\newcommand{\eq}[1]{(\ref{#1})}

\newcommand{\beq}{\begin{equation}}
\newcommand{\eeq}{\end{equation}}
\newcommand{\beqq}{\begin{equation*}}
\newcommand{\eeqq}{\end{equation*}}
\newcommand\beqa{\begin{eqnarray}}
\newcommand\eeqa{\end{eqnarray}}
\newcommand\beqaa{\begin{eqnarray*}}
\newcommand\eeqaa{\end{eqnarray*}}
\newcommand\bea{\begin{array}}
\newcommand\eea{\end{array}}

\def\tr{{\rm tr}\;}

\newcommand{\hs}{\frac{\sqrt{3}}{2}}
\renewcommand{\d}{\partial}

\newcommand{\<}{{\langle}}
\renewcommand{\>}{{\rangle}}

\newcommand{\cL}{{\cal L}}

\def\({\left(}
\def\){\right)}
\def\[{\left[}
\def\]{\right]}

\def\<{\langle}
\def\>{\rangle}

\def\s*{\ *_{\!\!\!\!\!\!\!\!\!\,_{\,_\text{\scriptsize{sym}}}}}
\def\hs*{\ \hat{*}_{\!\!\!\!\!\!\!\!\!\,_{\,_\text{\scriptsize{sym}}}}}
\def\d{\partial}

\begin{document}

\begin{flushright}
CERN-TH-2019-039
\end{flushright}
\vspace{-5truecm}
\title{The Holographic Fishchain}

\author{Nikolay Gromov}
\email{nikolay.gromov@kcl.ac.uk}
\affiliation{%
Mathematics Department, King's College London,
The Strand, London WC2R 2LS, UK
}%
\affiliation{St.Petersburg INP, Gatchina, 188 300, St.Petersburg,
Russia}
\vspace*{-250px}

\author{Amit Sever}%
\email{amit.sever@CERN.ch}
\affiliation{
School of Physics and Astronomy, Tel Aviv University, Ramat Aviv 69978, Israel}
\affiliation{
CERN, Theoretical Physics Department, 1211 Geneva 23, Switzerland
}%

\begin{abstract}
We present the first-principle derivation of a weak-strong duality between the fishnet theory in four dimensions and a discretized string-like model living in five dimensions. At strong coupling, the dual description becomes classical and we demonstrate explicitly the classical integrability of the model. We test our results by reproducing the strong coupling limit of the $4$-point correlator computed before non-perturbatively from the conformal partial wave expansion. Due to the extreme simplicity of our model, it could provide an ideal playground for holography with no super-symmetry. Furthermore, since the fishnet model and ${\cal N}=4$ SYM theory are continuously linked  our consideration could shed light on the derivation of AdS/CFT for the latter.

\end{abstract}

\pacs{Valid PACS appear here}
\maketitle

\section{\label{intro}Introduction}
In recent years the ideas of holography~\cite{Maldacena:1997re,Gubser:1998bc,Witten:1998qj} conquered almost all corners of theoretical physics. The idea that some (or any?) strongly coupled quantum system with many degrees of freedom should have an alternative dual description in terms of the gravity/string theory in a higher dimensional spacetime is becoming more and more popular. Despite this enormous attention the holographic principle has received in the last two decades, we are still lacking the first principle derivation of it. There are, however, numerous and extremely non-trivial tests of the duality. Due to its strong-weak character, it is very hard to produce these tests. 
In some special models such as ${\cal N}=4$ SYM theory, tools such as super-symmetric localization, or integrability, provide ways to compute observables for arbitrary coupling strengths and compare with the holographic predictions.

In this paper we will provide the first principle derivation of a holographic dual of the 
{\it fishnet model} \cite{Gurdogan:2015csr}
\beq\la{fishnet}
\cL_{4d} = N\,{\rm tr}\(|\d\phi_1|^2+|\d\phi_2|^2
+ (4\pi)^2 \xi^2 \phi_1^\dagger\phi_2^\dagger\phi_1\phi_2\)\,,
\eeq
in the planar expansion where $N$ is taken large while $\xi^2$ is held fixed but arbitrary \footnote{Here we have suppressed  double trace interactions which are not  relevant non-perturbatively \cite{Fokken:2013aea,Gromov:2018hut}.}. Here, $\phi_{1,2}$ are two $N\times N$ complex scalar fields. The model can be obtained from ${\cal N}=4$ SYM theory in a double scaling limit and was shown to be conformal and integrable in the planar limit \cite{Fokken:2013aea,Gurdogan:2015csr,Gromov:2017cja,Grabner:2017pgm}. 
This breaks the $\mathfrak{su}^*(4|4)$ superconformal symmetry down to $\mathfrak{u}(1)_1\times \mathfrak{u}(1)_2\times\mathfrak{so}(1,5)$ living us with no super-symmetry at all. Yet, being a solvable interacting CFT in four dimensions, this model attracted a lot of attention. In particular, one can compute the spectrum of anomalous dimensions \cite{Gromov:2017cja} as well as some structure constants and correlation functions \cite{Gromov:2018hut} at any $\xi$. In all of these cases, there were indications of the existence of the holographic dual -- the scaling dimensions $\Delta$ generically scale as $\xi$ and the $4$-point correlation functions behave as $e^{-\xi A(z,\bar z)}$ \cite{Gromov:2018hut}. At the same time, these indications were somehow puzzling as the dual description of ${\cal N}=4$ SYM become weakly coupled at infinitely large $\lambda$, whereas the fishnet model obtained in the
opposite $\lambda=R^2_{\rm AdS}/l^2_s\to 0$ limit \cite{Gurdogan:2015csr}. Furthermore, the corresponding deformation is known to produce tachionic instability in the string background \cite{Pomoni:2008de}. As a result, it is  not clear how to link the holographic string description of ${\cal N}=4$ SYM to that of the fishnets or even if that is possible at all.

Indeed, the dual description presented here is not in terms of a smooth string. Instead, we found a chain of $J$ particles or string-bits with nearest neighbor interactions. More precisely the dual model action functional $S_{\rm dual}=\xi\int dt\,\sum_iL_i$, is given in terms of the Lagrangian-density
\beq\la{actf}
L_i=-\frac{\dot X_i^2}{2}-\prod_{k=1}^J \(-X_k.X_{k+1}\)^{-\frac{1}{J}}-\eta_i (X_i^2+R^2)+R^2\;.
\eeq
Here, $X_i(t)\in{\mathbb R}^{1,5}$ with $-++\dots$ signature and $\eta_i(t)$, $R^2(t)$ are Lagrange multipliers. The world-sheet coordinates $X(t)$ are further subjected to Virasoro-type constraints described below in (\ref{Virasoro}) and (\ref{Vir2}). Note in particular that the square root of the 't Hooft coupling $\xi$, stands in front of the action and plays the role of $1/\hbar$.

The field $R(t)$ looks like an AdS radius in string units. It satisfies a dynamical evolution equation and will be set to zero consistently. The $X_i$ coordinates are not projective and hence, the action $S_{\rm dual}$ describe a discretized string propagating on the five dimensional lightcone of ${\mathbb R}^{1,5}$, subject to the Virasoro constraints. The fifth dimension naturally emerges when making the symmetries manifest. It is encoded in a non-trivial way in all the original $J$ four-dimensional degrees of freedom and is related to an emergent {\it local} scale invariance.

Below we give the derivation of this result and also show how one can reproduce the classical limit of the anomalous dimensions and also the 4-point functions.

\section{\label{fhshchain}Derivation of the dual action}
\begin{figure}
\centering
\includegraphics[scale=0.8]{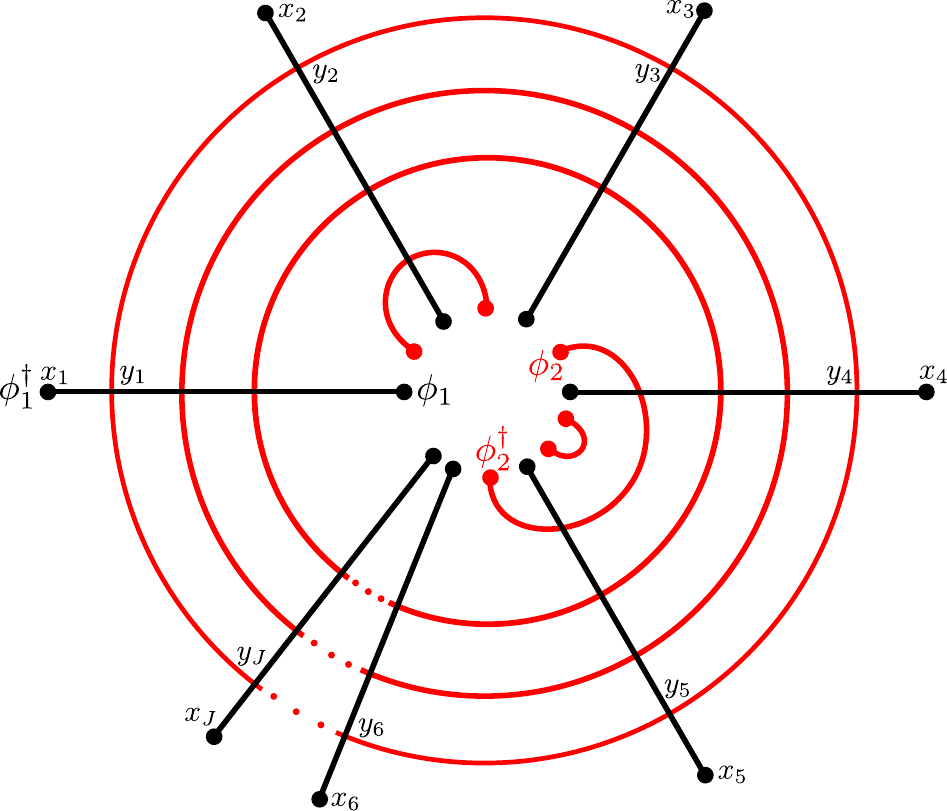}
\caption{Typical Feynman diagram in the $\mathfrak u(1)$ sector of the model has the wheel, or fishnet~\cite{Nielsen:1970bc},
structure which can be resummed. This structure also leads to integrability~\cite{Zamolodchikov:1980mb}.}
\label{fig:fishfig}
\end{figure}
One of the main features of the fishnet theory is the simple structure of its Feynman diagrams. In this paper we consider the ${\mathfrak u}(1)$ sector of the model, where the ${\mathfrak u}(1)_1$ charge is $J$ and the ${\mathfrak u}(1)_2$ charge is set to zero. It consists of all operators of the type $\tr(\d^m\phi_1^J(\phi_2\phi_2^\dagger)^n,\dots)$, containing any number of derivatives, $J$-scalar fields $\phi_1$ and any {\it neutral} combination of $\phi_2$ and $\phi_2^\dagger$. 

The Feynman diagrams which contribute to the correlation functions of these operators and their conjugates are of iterative fishnet type, after all $\phi_2$'s annihilate with $\phi_2^\dagger$, (see Fig.\ref{fig:fishfig}).
It is possible to resum, at least formally, infinitely many Feynman graphs by introducing the ``graph-building" operator $\widehat B$, defined by its integral kernel~\cite{Gurdogan:2015csr}
\beq\la{gb}
B(\{\vec y_i\}_{i=1}^J,\{\vec x_j\}_{j=1}^J)=\prod_{i=1}^J\frac{\xi^2/\pi^2}{(\vec y_i-\vec y_{i+1})^2(\vec x_i-\vec y_i)^2}\;.
\eeq
Applying  this operator once, we add one wheel to the graph on Fig.\ref{fig:fishfig}, thus the sum of all wheels inside the graph forms a geometric series
\beq
{\rm all\;wheels}=\frac{1}{1-\widehat B}\;.
\eeq
We see that the zeros of the denominator play a special role.
By diagonalizing $\widehat B$, one finds that the eigenfunctions are parameterized by the continuous parameter $\Delta$, conjugated to the dilatation operator. The sum over the complete set of eigenfunctions will involve the integration over $\Delta$, which then can be computed by residues giving distinct meaning to those values of $\Delta$ where $\widehat B=1$. Namely, those poles can be identified as the anomalous dimensions of the local operators. 
This procedure was exemplified in detail in \cite{Gromov:2018hut}. The output of this discussion is that we need to solve $(\widehat B-1)\Psi=0$, or, equivalently, acting on both sides with $\prod_i\Box_i$, to cancel factors $1/(4\pi^2(x_i-y_i)^2)$, we find
\beq\la{H1}
H\circ \Psi(\{x_i\})=0\ ,\quad {H}=\prod_{i=1}^J \vec p_i^2-\prod_{i=1}^J\frac{4\xi^2}{(\vec x_i-\vec x_{i+1})^2}
\eeq
where $\vec p_i\equiv -i \vec \partial_{x_i}$. Under the operator-state correspondence, the wave function $\Psi$ is dual to a local operator. The key step in our derivation is to interpret \eq{H1} as the constraint
appearing in a system with time reparametrization symmetry $t\to f(t)$, where $t$ is conjugate to $H$. To see this gauge symmetry manifestly, we write the Lagrangian corresponding to the Hamiltonian $H$ in \eq{H1}. After solving for $\vec p_i$ in term of $\dot {\vec x}_i=\frac{\d H}{\d \vec p_i}$ we arrive at 
\beq\la{Lpo}
{L}=\frac{2J-1}{2^{\frac{2J}{2J-1}}}\left(\frac{1}{\gamma}\prod_{i=1}^J \vec {\dot x}_i^2\right)^{\frac{1}{2 J - 1}}+\gamma\prod_{i=1}^J\frac{4\xi^2}{(\vec x_i-\vec x_{i+1})^2}
\eeq
where $\gamma$ transforms under reparametrization $t\to f(t)$ as $\gamma\to\gamma/ f'$ and (\ref{H1}) is the constraint that corresponds to fixing $\gamma=1$. Instead of fixing the gauge $\gamma=1$ it is more beneficial to eliminate the auxiliary field $\gamma$, by setting it to its extremum to obtain 
\beq\la{Sng}
S=\xi\int L\,dt=2 J\xi\int \(
\prod_{i=1}^J\frac{\dot {\vec x}_i^2}{(\vec x_i-\vec x_{i+1})^2}
\)^{\frac{1}{2J}}dt\;.
\eeq
One may draw analogies between \eq{Lpo}, \eq{Sng} and the Polyakov, Nambu-Goto actions respectively. 
In this analogy, the initial equation \eq{H1} corresponds to the Virasoro constraint. 
There is a number of significant observations one can make about \eq{Sng}. 
First, we see that the coupling $\xi$ is playing the role of $1/\hbar$ in the quasiclassical analysis in accordance with the previous observations~\cite{Gromov:2017cja,Gromov:2018hut}. In particular, that explains the scaling $\Delta\sim \xi$ observed numerically in \cite{Gromov:2017cja}.
We note, however, that our starting point \eq{gb}
contained $\xi^{2J}$, implying that all roots $ e^{\pi i n/J}\xi,\;n\in{\mathbb Z}$ should be considered. Different $n$'s correspond to different branches in the spectrum as we demonstrate in Sec.~\ref{examp}. 
In addition to the time reparametrization symmetry the action $S$ in the form \eq{Sng} is also invariant under global conformal transformations, which is of course a highly expected property for a CFT dual. In the next section, we will make the conformal symmetry manifest by uplifting the action into a 6D embedding space. 

\begin{figure}
\centering
\includegraphics[scale=0.34]{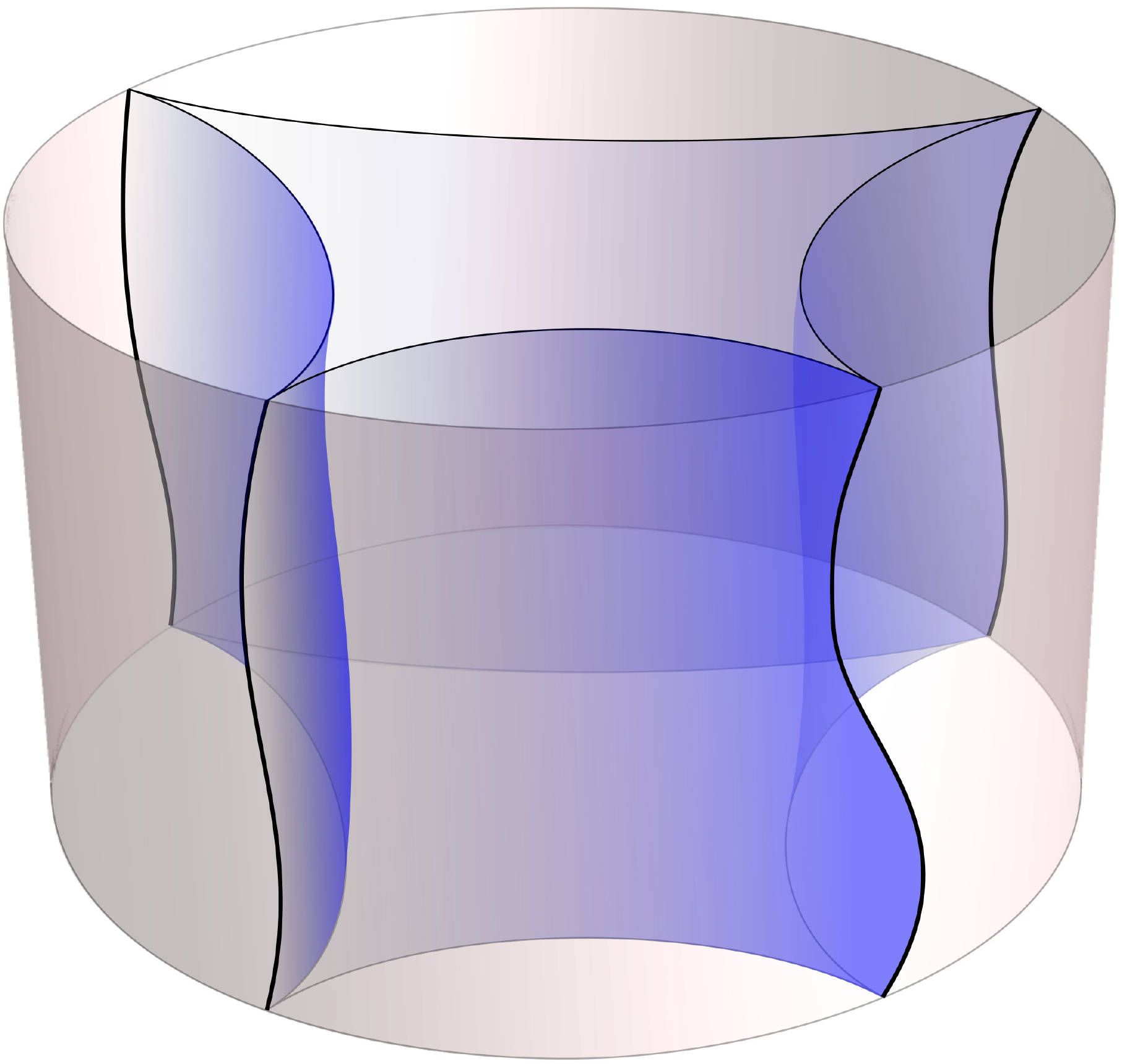}
\caption{Spiky string in AdS.}
\label{fig:fishfig3}
\end{figure}

Finally, we comment about an interpretation of the action (\ref{Sng}). One may think of it as describing $J$ string bits, \cite{Bergman:1995wh,Dorey:2008zy}. Each bit corresponds to a spike of a holographic string that propagates at the AdS boundary, see figure \ref{fig:fishfig3}. The dynamics of the string segment between the spikes results in the nearest neighbors interaction of the model.

\paragraph{Embedding space formulation.} As it is well known the conformal group in 4D coincides with the group of rotations in ${\mathbb R}^{1,5}$. Under this identification, the flat space coordinate $x^{\mu=1,\dots,4}$ is mapped to the projective lightcone of ${\mathbb R}^{1,5}$, parametrized by $X_i^M,\;M=-1,0,1,\dots,4$, as
\beq
x_i^\mu=X_i^\mu/X_i^+\ ,\quad X_i^2=0\ ,\quad X_i^+=X_i^0+X_i^{-1}\;.
\eeq
After this change of variables, the action (\ref{Sng}) becomes
\beq\la{Lup}
L=2 J\(
\prod_{i=1}^J\frac{\dot X_i\cdot \dot X_i}{-2X_i\cdot X_{i+1}}
\)^{\frac{1}{2J}}\;.
\eeq
By uplifting to a projective space, we have introduced a new local gauge symmetry $X_i\to g_i(t) X_i$. 

We now introduce auxiliary fields to disentangle the action as
\beq\la{actbetas}
L=-\sum_i 
\[\frac{\dot X_i^2}{2\alpha_i}+\eta_i X_i^2
+\gamma \prod_k \(-X_k.X_{k+1}\)^{-\frac{1}{J}}\]\;,
\eeq
where the last term is independent of $i$.
In order to get back to \eq{Lup} one should
extremize in $\alpha_i$ and $\gamma$, assuming $\prod \alpha_i=\gamma^J$.
We also introduced the remaining constraint $X_i^2=0$ with the Lagrange multipliers $\eta_i$. The symmetries of the action are 1) manifest conformal symmetry; 2) time-dependent re-scaling symmetry $X_i\to g_i(t)X_i,\;\alpha_i\to g_i^2(t)\alpha_i,\;\gamma\to \gamma \prod_i g_i^{2/J}(t)$ and $\eta_i\to g^{-2}_i(t)\eta_i$;
3) time reparameterization symmetry  $t\to f(t),\;\eta_i\to \eta_i/f',\;\gamma\to \gamma/f'$; 4) translation along the chain $X_{i}\to X_{i+1}$.
To fix the gauge symmetries we can set $\alpha_i=\gamma=1$, leading to the constraints
\beq\la{Virasoro}
\dot X_k^2=2\prod_i \(-X_i. X_{i+1}\)^{-\frac{1}{J}}\equiv {\cal L}\;,\;\;k=1,\dots,J\;,
\eeq
which is very reminiscent of the Virasoro constraints in the conformal gauge, telling us that the energy density is zero along the string. Finally, we notice that there is still one remaining gauge symmetry left $t\to f(t),\;X_i\to X_i/\sqrt{f'},\;\eta_i\to \eta_i/{f'}$,
which we fix by further imposing $\sum\eta_i=J$, with the
Lagrange multiplier $R^2$, leading to (\ref{actf}). This action together with \eq{Virasoro} is our main result~\footnote{alternative  gauge choice ${\cal L}=1$ could be convenient too}. \\

    \paragraph{Equations of motion.} The variation of \eq{actf} with respect to $X_i$ gives 
    \beq\la{eomX}
    \ddot X_i=2\eta_iX_i-{\cL\over2}\({X_{i+1}\over X_{i+1}.X_i}+{X_{i-1}\over X_i.X_{i-1}}\)\;.
    \eeq
    By contracting (\ref{eomX}) with $X_i$ and using that $X_i^2=0$ we arrive back at (\ref{Virasoro}). Contracting (\ref{eomX}) with $\dot X_i$ however, leads to the secondary constraint that is analogous of the second Virasoro constraint, imposing that 
    \beq\la{Vir2}
    {\dot X_i.X_{i+1}\over X_i.X_{i+1}}+{\dot X_i.X_{i-1}\over X_i.X_{i-1}}=-\d_t\log{\cL}
    \eeq 
    does not depend on the site index $i$, \footnote{The relative sign may look strange. However, in the continuum limit the analogue of the r.h.s. has the effect of correcting it~\cite{longpaper}.}.

    Finally, $\eta_i$ can be extracted from the derivative of (\ref{Vir2}). Instead, we  eliminate $\eta_i$ by introducing the $SO(1,5)$ charge density $q_i^{MN}=2\dot X_i^{[M}X_i^{N]}$. The equation of motion \eq{eomX} can be equivalently written as
    \beq\la{qcons}
    \dot { q}_i=\frac{\cL}2 ({ j}_{i+1}-{ j}_i)\ ,\quad { j}_i^{MN}=2{X_{i-1}^{[M}X_i^{N]}\over X_{i-1}.X_i}
    \eeq
    where $j_i$ can be interpreted an $SO(1,5)$ current density. The $SO(1,5)$ charge is given by $Q^{MN}\equiv \xi {\cal Q}^{MN}=\xi\sum_i{q_i^{MN}}$. 
    We can always assume that ${\cal Q}^{MN}$ is block diagonal, with non-zero elements
    ${\cal Q}^{-1,0}=i{\cal D}$, ${\cal Q}^{1,2}={\cal S}_1$ and ${\cal Q}^{3,4}={\cal S}_2$, where ${\cal D}=\Delta/\xi$ and ${\cal S}_a=S_a/\xi$ are the appropriate notations for the large $\xi$ classical limit. 
    
    After introducing $R^2$, we are no longer constrained to the lightcone and one could be worried about the consistency of the initial condition $X_i^2=-R^2=0$. By contracting \eq{eomX} with $X_i$ we get $2\d_t^2R^2=\eta_i R^2$. Since $\sum_i\eta_i=J$ we obtain $2\d_t^2R^2=R^2$, meaning that once we set $R=0$ at some moment of time it will stay so forever.

\paragraph{Integrability.} Our fishchain model at $R=0$ is dual to the integrable fishnet model and hence, it is expected to be integrable too. Similarly to the Toda chain, we find a pair of  spacelike and timelike connections, dependant on the spectral parameter $u$, ${\mathbb L}_i(u)$ and ${\mathbb V}_i(u)$, that satisfy the zero curvature condition \cite{book} 
\beq\la{zerocurvature}
\dot{\mathbb L}_i={\mathbb V}_{i+1}.{\mathbb L}_i-{\mathbb L}_i.{\mathbb V}_i\;.
\eeq
This condition ensures that each coefficient of the polynomial ${\mathbb T}(u)=\tr\Omega(u)$ where $\Omega\equiv{\mathbb L}_J\dots {\mathbb L}_2.{\mathbb L}_1$ gives an integral of motion, constant in time on equations of motions. In the irrep $\bf 6$ of $SO(1,5)$ these matrices are
\beq\la{LV6}
{\mathbb L}^{\bf 6}_i=u^2+uq_i+\frac{q_i^2}{2}\ ,\qquad{\mathbb V}^{\bf 6}_i={j_i\over u}\frac{\cL}{2}\;.
\eeq
To derive the discrete zero curvature condition (\ref{zerocurvature}) we use \eq{qcons} and the identity $\(q_i^2\)^{MN}=-\cL X_i^MX_i^N$,
which implies, using \eq{eomX}, that $\d_t q_i^2=\cL (j_{i+1}q_i-q_i j_i)$ and $j_{i+1}q_i^2-q_i^2 j_i=0$. Interestingly, the constraint \eq{Virasoro} results in the relation ${\mathbb T}^{\bf 6}(0)=(-1)^J$. Similarly to (\ref{LV6}), the spacelike and timelike connections in the irrep {\bf 4} take the form ${\mathbb L}^{\bf 4}_k=u-\frac{i}{2} q_k^{MN} \Sigma_{MN}$ and ${\mathbb V}^{\bf 4}_k=-\frac{i\cL}{4u}j_k^{MN}\Sigma_{MN}$, where $\Sigma_{MN}$ are the 6D $\sigma$-matrices. ${\mathbb L}^{\bf 6}$ can be constructed from ${\mathbb L}^{\bf 4}_k$ by projecting ${\mathbb L}^{\bf 4}_k\otimes{\mathbb L}^{\bf 4}_k$ on the {\bf 6}. 

The key objects in integrability are the $4$ quasi-momenta $p_a$ which are defined as $\det(\Omega^{\bf 4}(u)-u^J e^{ip_a(u)})=0$.
Their large $u$ asymptotic is determined by the global charges $p_a\simeq\frac{\pm \Delta\pm S_1\pm S_2}{2\xi u}$. At the origin $p_a$'s have a logarithmic singularity $\pm iJ\log u$. In addition, $p_a(u)$ has square-root singularities, coming from the diagonalization procedure.
Together they form an algebraic curve, whose genus depends on the number of degrees of freedom. We expect the number of cuts to be equal to the number of independent cross-ratios for $2J$ points (i.e. $2$ for $J=2$ and $8J-15$ for $J>2$). Each a-cycle on the curve corresponds to an {\it action} variable $I_a\equiv \frac{\xi}{\pi i}\oint_a p(u)\, du$. The action variables 
are expected to become integers in the Bohr-Sommerfeld quantization procedure. We postpone more detailed investigation of the algebraic curve and separation of variables in this model for the future~\cite{longpaper}.

\section{\label{test}Explicit example}\la{examp}
First we consider the simplest case where $J=2$. This case was studied in detail at the quantum level, in particular the spectrum is know exactly \cite{Grabner:2017pgm}
\beq\la{disp}
\Delta_{t=2/4}=2+
\sqrt{(S_1+1)^2+1\mp 2\sqrt{(S_1+1)^2+4\xi^4}}
\eeq
where $\pm$ correspond to the twist $t=2$ and twist $t=4$ branches in the spectrum. The $4$-point function was computed in \cite{Gromov:2018hut} as an infinite sum. In the classical limit, it was shown to sit on  saddle points with classical dimension and spins that are related to the two conformal cross ratios as
\beqa\la{data}
S_{\rm cl}^2=\pm\frac{4\xi^2\theta^2}{\theta^2+\rho^2}\quad\text{and}\quad\Delta_{\rm cl}^2=S_{\rm cl}^2\mp 4\xi^2\;,
\eeqa
where the second relation follows from \eq{disp}. Here, $\rho$ and $\theta$ parametrize the two conformal cross ratios $u=\frac{4}{(\cos \theta -\cosh \rho )^2}$ and $v=\frac{(\cos \theta +\cosh \rho )^2}{(\cos \theta -\cosh
\rho )^2}$. Furthermore, the 4-point correlation function itself takes the  form $e^{-\xi A_{\rm cl}}$ ($e^{i\xi A_{\rm cl}}$) for $t=2$ ($t=4$), with $A_{\rm cl}=2i\sqrt{\theta^2+\rho^2}$. Next, we try to reproduce this data from our classical dual description.

For $J=2$ we have two 6D null-vectors, $X_1(t)$ and $X_2(t)$. Using global symmetries we can always go to the centre of mass frame and set the last two components to zero
$
X_{1,2}=\frac{r}{\sqrt{2}}\(\cosh s,
\sinh s,\pm\cos\phi,\mp\sin\phi,0,0\)
$.
In this parametrization the coordinates $s$ and $\phi$ are conjugate to conserved charges, ${\cal D}=ir^2\dot s$ and ${\cal S}_1=r^2\dot\phi$. The constraint \eq{Virasoro} gives $\pm4=r^4(\dot s^2+\dot \phi^2)={\cal D}^2+{\cal S}_1^2$, where the $\pm$ sign comes from the different choice of the branch of the root in the r.h.s. of \eq{Virasoro}. It perfectly reproduces the spectrum \eq{disp}, \eq{data} in the classical limit where $S_1,\Delta\sim\xi\to\infty$, with different twists $t=2,4$ corresponding to different branches of the interaction term. The classical action, with the constraint taken into account, becomes $S=\mp4 \xi\int \frac{dt}{r^2}$. We see that it is beneficial to define the proper time $d\tau=\frac{dt}{r^2(t)}$, in terms of which 
\beq\la{SA}
S=\mp 4 \xi\,\tau\ ,\quad s=-i {\cal D}\,\tau\ ,\quad \phi={\cal S}_1\,\tau\;.
\eeq
Next, computing the cross-ratios between $X_{1,2}(\tau=0)$ and 
$X_{1,2}(\tau=T)$ we find in our parameterization $\theta = {\cal S}_1 T$ and $\rho=i{\cal D}\,T$. Next, solving for ${\cal S}_1$
and $T$, with the constraint ${\cal S}_1^2-{\cal D}^2=\pm 4$, we find $T^2=\pm\frac{1}{4}(\theta^2+\rho^2)$ and ${\cal S}_1^2=\pm\frac{4\theta^2}{\theta^2+\rho^2 },$ leading via \eq{SA} to $e^{iS}=e^{-2i\xi \sqrt{\theta^2+\rho^2 }}$ and $e^{iS}=e^{-2\xi \sqrt{\theta^2+\rho^2 }}$ for $t=2$ and $t=4$ correspondingly, in perfect agreement with  \cite{Gromov:2018hut}.

\begin{figure}
\centering
\includegraphics[scale=0.45]{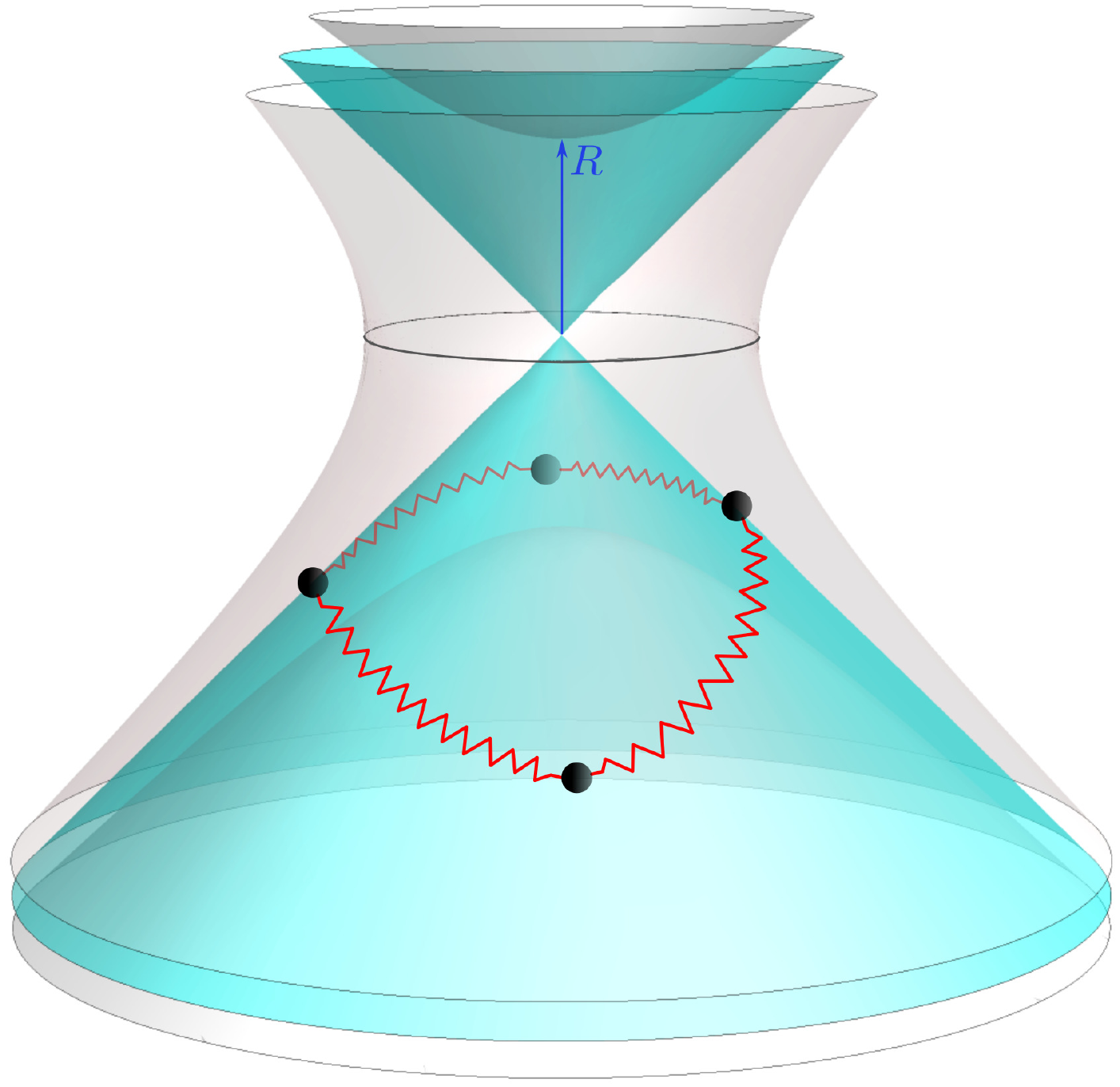}
\caption{Chain of $J$ particles on the light-cone of ${\mathbb R}^{1,5}$.}
\label{fig:cones}
\end{figure}

\section{Discussion and speculation}

There are two fundamentally important properties of the fishchain model (\ref{actf}). First, the square root of the 't Hooft coupling constant, $\xi$, stands in front of the action, playing the role of $1/\hbar$. It emerged naturally from our interpretation of the graph building operator. Second, the fishchain propagates in five-dimensional target space, see Fig.~\ref{fig:cones}. The fifth dimension has emerged from the principle of realizing all symmetries in a manifestly covariant way. It may be thought of as a concrete realization of the holographic map and the original prediction of 't Hooft \cite{tHooft:1973alw}.

Going away from the model (\ref{fishnet}), one may add back the rest of the fields of ${\cal N}=4$ SYM in a controlled expansion around the fishnet limit~\cite{Bykov:2012sc} and incorporate their effect on the dual fishchain. Such expansion may open the path for a rigorous proof of AdS/CFT. One way in which this path may materialize is the following. The radius $R^2$ came about as a Lagrange multiplier, associated with a global rescaling gauge symmetry. It is consistently set to zero in the relevant classical solutions discussed above. If the correction away from the fishnet limit will stabilize it at some fixed $R^2>0$ then the discretized string will propagate in AdS$_5$ instead of the lightcone of ${\mathbb R}^{1,5}$. Moreover, for fixed $R^2>0$, the model has a smooth large $J$ continuum limit where $X_i.(X_{i+1}-X_i)\simeq {1\over2}\epsilon^2X''$. As a result, a new local time reparametrization symmetry emerges. Using similar manipulations to the ones in section \ref{fhshchain} one arrives at a string action in AdS$_5$ that is subject to the standard local Virasoro constraint, including a constant contribution from the extension of the string on the sphere \cite{longpaper}. In the $J\to\infty$ limit we may also be able to find a connection with the proposal of \cite{Basso:2018agi}. 

There are many future directions to pursue, some of them we list below. ${\mathrm I}$) One would like to extend the fishchain model away from the ${\mathfrak u}(1)$ sector by incorporating extra $\phi_2$ fields. $\mathrm{II}$) One may incorporate quasi-classical corrections. $\mathrm{III}$) Systematic $1/N$ expansion should lead to fishchain interaction vertices. $\mathrm{IV}$) The open fishchain version of the model, dual to Wilson lines in the ladder limit~\cite{Correa:2012nk}, can be obtained from the derivation above by adding two more sites, replacing $\vec p^2_{1,J+2}\to\vec p^2_{1,J+2}+m^2$ and taking the large mass limit. $\mathrm{V}$) We expect the fishchain to exhibit T-duality. $\mathrm{VI}$) The simplicity of the classical model could help with the separation of variables approach~\cite{Cavaglia:2018lxi,twistingpaper} to the correlation functions in ${\cal N}=4$ SYM.

Finally, analogous fishnet diagrams also exist in 2,3 and 6 dimensions \cite{Zamolodchikov:1980mb} and one may try to derive their duals. In particular, one may consider the large twist limit of the ABJM model \cite{Aharony:2008ug,Caetano:2016ydc}, or more general fishnets, which could also include fermions~\cite{Kazakov:2018gcy}.

{\it Acknowledgements:} We thank 
D.~Anninos, B.~Basso, A.~Cavagli\`a,
G.~Korchemsky, F.~Levkovich-Maslyuk,
I.~Kostov, S.~Lukyanov, F.~Smirnov, 
K.~Zarembo,
B.~Vicedo,
S.~Zhiboedov and especially to V. Kazakov for invaluable discussions. N.G. acknowledges the hospitality of the theory group at CERN during the main phase of this work.
N.G. was supported by the STFC grant (ST/P000258/1). A.S. was supported by the I-CORE Program of the Planning and Budgeting Committee, The Israel Science Foundation (1937/12) and by the Israel Science Foundation (grant number 968/15).

\end{document}